\documentstyle[12pt]{article}

\newcommand{\EQ}{\begin{equation}}
\newcommand{\EN}{\end{equation}}
\newcommand{\s}{\sigma}
\newcommand{\goto}{\rightarrow}
\begin{document}
\setcounter{page}{0}
\topmargin 0pt
\oddsidemargin 5mm
\renewcommand{\thefootnote}{\arabic{footnote}}
\newpage
\setcounter{page}{0}
\begin{titlepage}
\begin{flushright}
Swansea SWAT/94-95/64
\end{flushright}
\vspace{0.5cm}
\begin{center}
{\large {\bf Random Bond Ising Model and Massless Phase of the Gross-Neveu
Model}} \\
\vspace{1.8cm}
{\large G. Mussardo\footnote{Permanent address:
International School for Advanced Studies and
Istituto Nazionale di Fisica Nucleare, Trieste, Italy.}
and P. Simonetti}\\
\vspace{0.5cm}
{\em Department of Physics and Mathematics,\\
University College of Swansea, \\
Swansea SA2 8PP, UK}\\

\end{center}
\vspace{1.2cm}

\renewcommand{\thefootnote}{\arabic{footnote}}
\setcounter{footnote}{0}

\begin{abstract}
\noindent
The $O(n)$ Gross-Neveu model for $n < 2$ presents a massless phase that can
be characterized by right-left mover scattering processes. The limit
$n \goto 0$ describes the on-shell properties of the random bond Ising model.
\end{abstract}

\vspace{.3cm}

\end{titlepage}

\newpage
%\resection{Introduction}

\noindent
1. Aim of this paper is to discuss the $S$-matrix formulation of the
$O(n)$ Gross-Neveu (GN) model for $n<2$, and in particular the scattering
theory associated to the limit $n \rightarrow 0$. There is a well-defined
physical problem related to this limit, which is the analysis of the influence
of random impurities, induced by thermal fluctuations, on the critical
behaviour of pure homogeneous systems. It is therefore useful to briefly
remind the relationship between the GN model and the random systems.

For  ``random temperature" kinds of impurities, i.e. defects or dislocations
in the material induced by thermal fluctuations,  there is a simple criterion
\cite{harris} for extimating the relevance of weak disorder on a critical
system. According to this criterion, the effect of the disorder depends on
the sign of the specific heat critical index $\alpha$ of the pure material.
For $\alpha>0$ the impurities are expected to completely suppress the long
range fluctuations of the pure system, canceling all singularities in the
thermodynamical quantities. On the contrary, for $\alpha < 0$, the impurities
may produce a shift of the critical temperature but they do not affect the
critical behaviour, i.e. the critical exponents are the same as in the pure
system. The marginal case $\alpha=0$ is special and must be separately
analysed. This situation occurs in the two-dimensional Ising model with
random bond distribution, and it has been initially considered by Dotsenko
and Dotsenko \cite{dotdot}. In particular, these authors have shown that near
the critical temperature, the class of universality of the random bond Ising
model is described by the $O(n)$ Gross-Neveu model in the limit $n\goto 0$
(see also \cite{shalaev,shankar,ludwig}). The mapping is realized as follows.
The two-dimensional homogeneous Ising model can be described in the continuum
limit by a massive Majorana fermion $\Psi$, where the mass $m$ is a linear
measurament of the deviation of the temperature $T$ from the critical
value $T_c$. The partition function for the pure system is given by
\EQ
Z[m] = \int [d\Psi] \exp\left[-\int d^2x \overline \Psi (i \not\hspace{-0.07cm}
\partial - m) \Psi
\right] \,\,\, .
\EN
Suppose now that the temperature, instead of being uniform on the
whole lattice, is allowed to vary from point to point but staying close,
in average, to the critical value. The mass parameter becomes then a
random variable $m(x)$, with a probability distribution $P[m(x)]$
which is supposed to be gaussian, $P[m(x)]\sim \exp\left[-m^2(x)/2 g^2\right]$.
Assuming that the time scale of the spin-flips is much faster than the
typical time scale in which the local temperatures are updated, one is
led to consider the quenched average on the disorder variables, i.e.
the average of the free energy rather than the average of the partition
function
\EQ
\overline{\ln Z[m(x)]} = \int \ln Z[m(x)]\, P[m(x)]\, dm(x) \,\,\, .
\label{quen}
\EN
A standard method to get around the difficulty presented by this computation
is to use the replica trick \cite{rt}, which consists in replacing
$$
\ln Z \goto \lim_{n\goto 0} \frac{Z^n -1}{n}
$$
in the integral of the right hand side of (\ref{quen}). In this way, taking
the quenched average for the random bond Ising model is equivalent to solve
the dynamics of the Gross-Neveu model in the limit $n \goto 0$
\EQ
\overline{\ln Z} \sim  \int [d\Psi_i] \exp\left[-{\cal S}\right]
\,\, ,
\EN
where
\EQ
{\cal S}\, =\, \int d^2 x
\left[\overline \Psi_i \not\hspace{-0.07cm}\partial \Psi_i + \frac{g^2}{2}
(\overline \Psi_i \Psi_i)^2 \right] \,\,\,\,\,, \,\,\, i=1,\ldots,n.
\label{lagr}
\EN

\vspace{5mm}

\noindent
2. It is well known that the Gross-Neveu model \cite{GN} presents different
properties for $n >2$ and for $n < 2$ ( for $n=2$ it coincides with
the massless Thirring model, which is a Conformal Field Theory). This can be
seen, for instance, by considering the $\beta$ function of the theory, that
at the first order reads \cite{wetzel}
\EQ
M\frac{d}{d M} g(M) = \beta(g(M)) = -(n-2) \frac{g^3}{2\pi} + \cdots
\label{rg}
\EN
($M$ is a mass scale). For $n>2$ the theory is asymptotically free in the
ultraviolet limit and there is a dynamical mass generation. The chirality
symmetry of the original lagrangian (\ref{lagr}) is spontaneously broken and
the fermionic fields are actually massive. The resulting theory is integrable
and the exact elastic $S$-matrix has been computed in \cite{zz,kt}. In
addition to the elementary fermions appearing in the lagrangian (\ref{lagr}),
the final spectrum of the theory comes out to have quite a rich structure of
bound states.

The situation is rather different when $n < 2$. Since the $\beta$ function
is now positive around the origin, the GN model becomes an infrared
asymptotically free theory and therefore a dynamical mass generation
for the physical particles cannot occur \cite{GN}. In this case, we can
extract the spectrum by directly looking at the infrared limit of the
lagrangian (\ref{lagr}): it simply consists in $n$ copies of massless
Majorana fermions and no additional low-energy excitations are expected.
Since the $\beta$-function is not zero, the model is massless but not
conformally invariant and therefore the mass scale $M$ entering the
Renormalization Group equation (\ref{rg}) characterises the cross-over of the
theory going from the infrared to the ultraviolet region. This massless phase
of the GN model consists then of $n$ right-moving particles $R_a(p)$ with the
energy spectrum $E = p$ ($p > 0$) and $n$ left-moving particles $L_a(p)$
($ p < 0$) with $E = -p$. These dispersion relations can be conveniently
parameterized in terms of the mass scale $M$ and the rapidity variable
$\theta$ as follows: $E = p = \frac{M}{2} e^{\theta}$ for the right-movers
and $E = -p = \frac{M}{2} e^{-\theta}$ for the left-movers.

\vspace{5mm}

\noindent
3. Assuming that the integrability of the GN model also holds for $n < 2$, we
can compute the exact elastic $S$-matrix associated to this massless phase
of the model (for the definition of massless scattering and its properties,
we refer the reader to the original references \cite{ml1,ml2}). In the
scattering processes of integrable massless theories, the numbers of left
and right particles are separately conserved and there is a factorization of
the amplitudes. Hence we can restrict our attention only to the two-body
processes. For the infrared asymptotic freedom of the model, we expect no
scattering of the left-left or right-right particles. Therefore, in these two
sectors the $S$-matrix is simply $-1$. On the contrary, in the right-left
sector, the $S$-matrix is defined by the commutation relation
\EQ
R_a(\theta_1) L_b(\theta_2) =  S_{ab}^{cd}(\theta_1 - \theta_2) L_d(\theta_2)
R_c(\theta_1) \,\,\, .
\label{algebra}
\EN
The $S$-matrix has to respect the $O(n)$ invariance of the Lagrangian
(\ref{lagr}) and can consequently be decomposed as
\EQ
S_{a,b}^{c,d}(\theta)\, = \,\delta_{a b} \delta^{c d} \s_1(\theta) +
\delta_a^c \delta_b^d\, \s_2(\theta) + \delta_a^d \delta_b^c \,
\s_3(\theta) \,\,\, .
\EN
Equivalently, we may consider the $S$-matrix for the channels of definite
isospin,
\begin{eqnarray}
&& \sigma_{isoscalar} = N \sigma_1 + \sigma_2 + \sigma_3 \,\,\,;\nonumber \\
&& \sigma_{antisym} = \sigma_2 - \sigma_3 \,\,\,;
\label{iso} \\
&& \sigma_{sym} = \sigma_2 + \sigma_3 \,\,\,.\nonumber
\end{eqnarray}
The above amplitudes have to satisfy the unitarity and crossing symmetry
equations, which for the massless scattering are given respectively by
\EQ
\begin{array}{c}
\sigma_2(\theta) \sigma_2^*(\theta) + \sigma_3(\theta)
\sigma_3^*(\theta) = 1 \,\,\,;
\\
n\,\sigma_1(\theta) \sigma_1^*(\theta) + \sigma_1(\theta)
\sigma_2^*(\theta) +
\sigma_1(\theta) \sigma_3^*(\theta) + \sigma_2(\theta)
\sigma_1^*(\theta) +
\sigma_3(\theta) \sigma_1^*(\theta) = 0 \,\,\,;
\\
\sigma_2(\theta) \sigma_3^*(\theta) + \sigma_3(\theta) \sigma_2^*(\theta)
= 0 \,\, ,
\end{array}
\label{un}
\EN
and
\EQ
\begin{array}{ll}
\sigma_1^*(\theta) & = \sigma_3(i \pi + \theta) \,\,\,; \\
\sigma_2^*(\theta) & = \sigma_2(i \pi + \theta) \nonumber \,\,\,.
\end{array}
\label{cr}
\EN
The associativity condition for the algebra (\ref{algebra}) gives rise to
the Yang-Baxter equations, whose solution is given by
\begin{eqnarray}
&&
\sigma_1(\theta) = - \frac{i \lambda}{i \lambda \left(\frac{n-2}{2}\right)
- \theta} \,\sigma_2(\theta) \,\,\,; \label{yb}
\\
&& \sigma_3(\theta) = - \frac{i\lambda}{\theta} \sigma_2(\theta) \,\, ,
\nonumber
\end{eqnarray}
with the parameter $\lambda$ fixed by the crossing symmetry equations
(\ref{cr}) to be
\EQ
\lambda \,= \,\frac{2\pi}{n-2} \,\,\, .
\EN
Inserting (\ref{yb}) into eq.\,(\ref{un}) and using (\ref{cr}), we arrive to
the combined unitarity-crossing equation satisfied by $\sigma_2$,
\EQ
\sigma_2(\theta) \sigma_2(i \pi + \theta) \,=\, \frac{\theta^2}{\theta^2 +
\lambda^2}
\,\,\, .
\EN
The minimal solution is given by
\EQ
\sigma_2(\theta) \,=\,  -\,
\frac{\Gamma\left(1 - \frac{\theta}{2\pi i}\right)
\Gamma\left(\frac{1}{2} + \frac{\theta}{2\pi i}\right)
\Gamma\left(\frac{1}{2} - \frac{\lambda}{2 \pi} - \frac{\theta}{2\pi i}\right)
\Gamma\left(- \frac{\lambda}{2 \pi} + \frac{\theta}{2\pi i}\right)}
{\Gamma\left(\frac{\theta}{2\pi i}\right)
\Gamma\left(\frac{1}{2} - \frac{\lambda}{2\pi} + \frac{\theta}{2\pi i}\right)
\Gamma\left(\frac{1}{2} - \frac{\theta}{2\pi i}\right)
\Gamma\left(1 - \frac{\lambda}{2\pi} - \frac{\theta}{2 \pi i}\right)}
\,\,\, .
\label{smatr}
\EN
This expression coincides with the analogous formula obtained in the GN model
for $n > 2$ but there is an important difference here: for $n < 2$,
$\lambda$ is negative and therefore the amplitudes present no poles in the
physical strip. The absence of poles in the physical strip  matches of course
with the nature of massless scattering. It is easy to see that for $n=1$
(which is simply the massless Ising model) the isoscalar $S$-matrix in
(\ref{iso}) correctly reduces to $-1$ whereas for $n \goto 0$, the relevant
limit for the random bond Ising model, we have
\EQ
\sigma_2(\theta) \,=\, -\,\frac{\theta}{(i\pi-\theta)}\,
\left( \frac{
\Gamma\left(\frac{1}{2} + \frac{\theta}{2\pi i}\right)
\Gamma\left(- \frac{\theta}{2\pi i}\right)}
{\Gamma\left(\frac{1}{2} - \frac{\theta}{2\pi i}\right)
\Gamma\left(\frac{\theta}{2\pi i}\right)}  \right)^2 \,\,\,.
\label{smatr0}
\EN
In conclusion, our analysis shows that for $n < 2$ the $S$-matrix of the
elementary fermions of the GN model may be regarded as the analytic
continuation of the $S$-matrix for $n > 2$ and viceversa. The interpretation,
however, is quite different in the two intervals. In fact, for $n > 2$ the
$S$-matrix describes the interaction of massive fermions and, in order to
close the bootstrap, has to be supported by the computation of the other
scattering amplitudes relative to the bound states \cite{zz,kt}. For $n < 2$,
on the contrary, there are no bound states, the fermionic fields split into
the chiral components and the $S$-matrix describes in this case only the
interaction of the left-right movers\footnote{Obviously, also the
interpretation of the rapidity variable changes in the two intervals.}.

\vspace{5mm}

\noindent
4. The massless $S$-matrix proposed in this paper leaves open some important
questions. The first question concerns the characterization of the ultraviolet
limit of the GN model for $n < 2$. In fact, the perturbative approach to an
infrared free theory is usually plagued by the presence of Landau pole which
may prevent to extend the validity of the theory beyond it. If this is the
case, the ultraviolet region cannot be reached perturbatively from the
infrared fixed point. The massless GN $S$-matrix (\ref{smatr}) has no
singularities for real values of $\theta$. This seems to indicate that for
the massless GN model, the ultraviolet limit is, instead, continuosly
connected to the infrared fixed point. In order to clarify this point, a first
step would be to analyse the behaviour of the massless GN model on an
infinite cylinder of width $R$ and extract, by means of the Thermodynamical
Bethe Ansatz, the flow of the effective central charge $\tilde c(MR)$ between
the ultraviolet ($ MR \goto 0$) and the infrared ($MR \goto \infty$) fixed
points. The second question concerns the local properties of the GN model
and their relationship with the original statistical system described by the
limit $n \goto 0$. As for the limit $n \goto 0$ of the bosonic $O(n)$
model studied in \cite{zpol,cm} in connection to the self-avoiding walks, it
would be interesting to apply the massless Form Factor approach, recently
considered in \cite{dms}, and study the behaviour of the correlation functions
of the random bond Ising model associated to the $S$-matrix
of eq.\,(\ref{smatr0}).

\vspace{0.5cm}
\noindent
{\em Acknowledgements.}
\noindent
We are grateful to G. Delfino, J.L. Cardy, T. Hollowood and A. Schwimmer
for useful discussions. One of us (GM) thanks Prof. D.I. Olive and the
Theoretical Physics Department of University College of Swansea for warm
hospitality.

\end{document}